\documentclass[prb,reprint,noshowkeys,showpacs,amssymb,amsmath,amstext,amsfont]{revtex4-1}
\usepackage{graphicx}
\usepackage{bm}%

\newcommand{\SSS}{\scriptscriptstyle}

\newcommand{\Ii}{{\rm i}}

\newcommand{\RE}{{\mathrm{Re}}}
\newcommand{\IM}{{\mathrm{Im}}}

\newcommand{\KapO}{\kappa_{\mathrm{o}}}
\newcommand{\KapE}{\kappa_{\mathrm{e}}}

\newcommand{\EpsE}{\epsilon_{\mathrm{e}}}
\newcommand{\EpsO}{\epsilon_{\mathrm{o}}}

\newcommand{\SigG}{\sigma_{\mathrm{g}}}
\newcommand{\EF}{E_{\SSS\mathrm{F}}}
\newcommand{\VF}{v_{\SSS\mathrm{F}}}

\newcommand{\OmgTE}{\omega_{{\SSS\mathrm{T}}\mathrm{e}}}
\newcommand{\OmgTO}{\omega_{{\SSS\mathrm{T}}\mathrm{o}}}
\newcommand{\OmgLE}{\omega_{{\SSS\mathrm{L}}\mathrm{e}}}
\newcommand{\OmgLO}{\omega_{{\SSS\mathrm{L}}\mathrm{o}}}

\newcommand{\GamG}{\gamma_{\mathrm{g}}}

\begin{document}

\title{Ferroelectric-Gated Terahertz Plasmonics on Graphene}
\author{Dafei Jin}
\author{Anshuman Kumar}
\author{Kin Hung Fung}
\author{Jun Xu}
\author{Nicholas X. Fang}\email{nicfang@mit.edu}\affiliation{Department of Mechanical Engineering,
Massachusetts Institute of Technology, Cambridge, Massachusetts 02139, USA}
\date{\today}

\begin{abstract}
Inspired by recent advancement of low-power ferroelectic-gated memories and transistors, we propose
a design of ferroelectic-gated nanoplasmonic devices based on graphene sheets clamped in
ferroelectric crystals. We show that the two-dimensional plasmons in graphene strongly couple with
the phonon-polaritons in ferroelectrics at terahertz frequencies, leading to characteristic modal
wavelength of the order of 100--200~nm at only 3--4~THz. By patterning the ferroelectrics into
different domains, one can produce compact on-chip plasmonic waveguides, which exhibit negligible
crosstalk even at 50~nm separation distance. Harnessing the memory effect of ferroelectrics,
low-power electro-optical switching can be achieved on these plasmonic waveguides.

\end{abstract}


\maketitle

The emergence of graphene research in the recent years has triggered a significant interest in
two-dimensional
plasmonics.\cite{JablanPRB2009,LongNatureNano2011,RyuACSNano2011,KoppensNanoLett2011,DavoyanPRL2012,BaoACSNano2012,VakilScience2011,ChenNature2012,FeiNature2012}
The transport behaviors of graphene reveal extremely low ohmic loss and nearly perfect
electron-hole
symmetry.\cite{GeimNatureMaterials2007,BolotinPRL2008,MorozovPRL2008,PeresRMP2010,DasSarmaRMP2011}
The charge-carrier density (or the Fermi level) can be conveniently adjusted via chemical doping
and electrostatic gating, which give rise to tunable plasmonic oscillations in the terahertz
frequency regime.\cite{LongNatureNano2011,RyuACSNano2011} In combination with its remarkable
character of single-atom thickness and the ability of subwavelength light confinement, graphene has
become a promising platform for the new-generation nanoplasmonic
devices.\cite{KoppensNanoLett2011,BaoACSNano2012} So far the most widely studied graphene plasmonic
structures are fabricated on silicon dioxide (SiO$_2$) substrates or suspended in
air,\cite{ChenNatureNano2008,NikitinPRB2011,ChristensenACSNano2012} where the SiO$_2$ and air serve
as the dielectric claddings for the creation of surface plasmon-polaritons
(SPPs).\cite{EconomouPR1969} In contrast with conventional dielectrics, ferroelectrics such as
lithium niobate (LiNbO$_3$) and lithium tantalate (LiTaO$_3$) bear giant permittivity and
birefringence at terahertz frequencies.\cite{FeurerARMR2007,SunCPL2007} This peculiarity is
associated with the coexistence of terahertz phonon-polariton
modes\cite{AshcroftSolidStatePhysics1976,MarekPhDThesis2003} and static macroscopic
polarization.\cite{SetterJAP2006,DawberRMP2005,SannaPRB2010} It is thus of particular interest to
investigate whether the unique properties of ferroelectrics can be integrated into
nanoplasmonics,\cite{LiuAPL2006,SpanierNanoLett2006,DickenNanoLett2008} and especially, display a
strong coupling with graphene plasmonics in the same terahertz range.

\begin{figure}
\centerline{\includegraphics[scale=0.7]{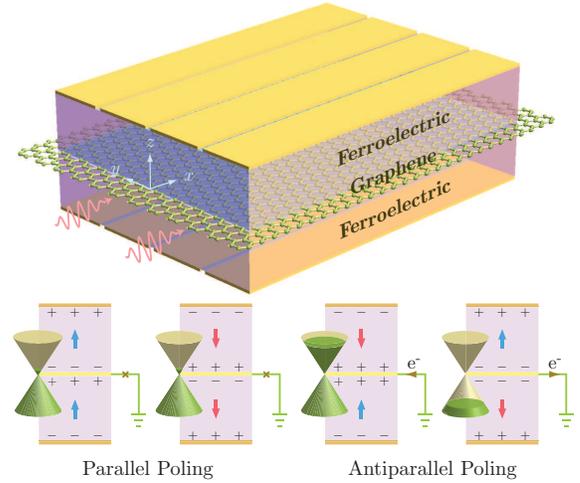}} \caption{Schematics of a
ferroelectric-graphene-ferroelectric structure with parallel- and antiparallel-poling domains of
the macroscopic polarization in the ferroelectrics. Parallel poling does not change the Fermi level
of graphene from (possible) chemical doping, while antiparallel poling can induce an extra
(positive or negative) electrostatic gating effectively.}\label{FigDesign}
\end{figure}

In this paper, we propose a design of ferroelectic-gated graphene plasmonic devices operating at
THz frequencies. Figure~\ref{FigDesign} illustrates the building blocks, while more complex
structures can be made based on the same idea. This type of architecture has at least two appealing
features that are not present in the common dielectric-graphene-dielectric architecture, and may
bring fresh insight into the design of (classical or quantum) optical circuits. The first feature
originates from the extremely large permittivity and comparatively high $Q$-value near the
optical-phonon resonances of ferroelectrics at terahertz frequencies.\cite{FeurerARMR2007} When the
two-dimensional plasmons in graphene become coupled with the phonon-polaritons in ferroelectrics by
exchanging photons across their interfaces, they form the so-called surface
plasmon-phonon-polaritons (SPPPs) and lead to around 100~nm modal wavelength, even if the driving
frequency remains at a few THz. These subwavelength modes are fundamentally supported by the
atomic-level oscillations of ferroelectric ions and are limited by the dissipation through
anharmonic phonon processes.\cite{AshcroftSolidStatePhysics1976,HarhiraPSSC2007,ArakelianPSSB1991}
The second feature lies in the large macroscopic polarization in ferroelectrics of the order of
$10$~$\mu$C~cm$^{-2}$, corresponding to a large surface bound-charge density of the order of
$10^{-14}$~cm$^{-2}$.\cite{DawberRMP2005,SetterJAP2006,MarekPhDThesis2003} The parallel- and
antiparallel-poling configurations can induce drastically different free-charge densities on
graphene, which effectively switch off and on the Fermi energy by about $1$~eV. These features may
be employed for constructing nanoplasmonic elements (waveguides, resonators, antennae, etc.)
through the electrostatic gating of ferroelectrics.\cite{HaussmannNanoLett2009,LiARMR2008} Thanks
to the memory effect of ferroelectrics, an intended design can sustain itself without a need of
constant input bias, and can be conveniently refreshed in a low-power manner, similar to those
operations in the ferroelectric random-access memories (FeRAM) and ferroelectric field-effect
transistors (FeFET).\cite{DawberRMP2005,SetterJAP2006,ZhengAPL2009,SongAPL2011}

Let us first study the eigen-modes on an infinite ferroelectric-graphene-ferroelectric structure.
Suppose the materials being stacked along the $z$-axis, and infinitely extended in the $xy$-plane,
matching the same coordinate system indicated in Fig.~\ref{FigDesign}. The graphene sheet is
situated at $z=0$ with a zero thickness yet a nonzero two-dimensional conductivity
$\SigG$.\cite{JablanPRB2009,KoppensNanoLett2011,BaoACSNano2012} The two ferroelectric crystals
occupy the semi-infinite regions $z<0$ and $z>0$, respectively. We assume the macroscopic
polarizations of the ferroelectrics to be aligned with either $+z$ or $-z$ direction, so the
optical axes of the crystals always coincide with the
$z$-axis.\cite{MarekPhDThesis2003,SannaPRB2010} The static polarity mainly affects the electron
density on graphene but has no appreciable impact on the optical properties in bulk
crystals.\cite{DawberRMP2005,SetterJAP2006,ZhengAPL2009,SongAPL2011} The eigen-solutions of the
entire structure can be labeled by frequency $\omega$ and in-plane wavenumbers $k_x$ and $k_y$
across all the regions. Within each region, the electric field $\bm{E}$ and the magnetic field
$\bm{H}$ are linear combinations of plane waves associated with an out-of-plane wavenumber $k_z$.
For surface-wave solutions, $k_z$ is an imaginary number. Due to anisotropy, we may define the
ordinary and extraordinary two evanescent wavenumbers with respect to the optical $z$-axis, $
\KapO^2 \equiv -k_z^2 = (k_x^2 + k_y^2) - (\omega^2/c^2)\EpsO $, $\KapE^2 \equiv -k_z^2 = (k_x^2 +
k_y^2) \EpsO/\EpsE - (\omega^2/c^2)\EpsO$, respectively. $\EpsO$ and $\EpsE$ are the ordinary and
extraordinary permittivities of the ferroelectrics.\cite{FeurerARMR2007,SunCPL2007} After some
boundary treatment, we shall find the dispersion relation (in the cgs units),
\begin{equation}
2\frac{\EpsO}{\KapE} = \frac{4\pi \SigG}{\Ii \omega}, \label{Dispersion}
\end{equation}
which is an anisotropic generalization to the dispersion relation in dielectric-graphene-dielectric
structures.\cite{JablanPRB2009}

In graphene plasmonics, the Fermi energy $\EF$ measured from the Dirac point usually ranges from
about $\pm 0.05$~eV to $\pm1.0$~eV, equivalent to an electron or hole concentration
$n_{\mathrm{c}}$ varying from $1.84\times10^{11}$~cm$^{-2}$ to $7.35\times10^{13}$~cm$^{-2}$
according to the relation $n_{\mathrm{c}}=(|\EF|/\hbar \VF)^2/\pi $, where $\VF$ is the Fermi
velocity taken as $1\times
10^8$~cm~s$^{-1}$.\cite{GeimNatureMaterials2007,BolotinPRL2008,MorozovPRL2008,PeresRMP2010,DasSarmaRMP2011}
For the experiments below 10~THz, the simple Drude formula can quite accurately describe the
graphene conductivity (in the cgs
units),\cite{JablanPRB2009,LongNatureNano2011,KoppensNanoLett2011,DavoyanPRL2012,BaoACSNano2012}
$\SigG = \Ii e^2|\EF|/\pi\hbar^2(\omega + \Ii \GamG)$, where $\GamG$ is the relaxation rate. For an
ultra-pure sample, the usual mobility limitation comes from the electron scatterings with the
thermally-excited acoustic phonons in graphene and the remote optical phonons from
SiO$_2$.\cite{BolotinPRL2008,MorozovPRL2008,ChenNatureNano2008,PeresRMP2010,DasSarmaRMP2011}
According to the literature, we estimate the mobility for our structure to be of the order of
$1\times 10^5$~cm$^2$~V$^{-1}$s$^{-1}$ at room temperature ($\sim300$~K), and
$1\times10^6$~cm$^2$~V$^{-1}$~s$^{-1}$ at low temperature ($\sim100$~K).\cite{HongPRL2009} The
relaxation rate can be calculated from mobility via $\GamG =
e\VF^2/\mu|\EF|$.\cite{PeresRMP2010,DasSarmaRMP2011}

We choose LiNbO$_3$ as an example of ferroelectric materials. It is especially convenient for
low-THz experiments, because the extraordinarily polarized THz waves can be triggered by 800~nm
femtosecond laser pulses via nonlinear optical response inside
LiNbO$_3$.\cite{FeurerARMR2007,HeblingIEEEJQE2008} According to the Raman scattering
data,\cite{MarekPhDThesis2003,IvanovaFerro1978,CapekPSSC2007,HuaMCP2002} LiNbO$_3$ has two
fundamental transverse optical-phonon frequencies $\OmgTO$ and $\OmgTE$, corresponding to the
ordinary and extraordinary waves, respectively. The main behaviors of the permittivities $\EpsO$
and $\EpsE$ in the 0.1--10~THz frequency range can be fitted by the Lorentz
model,\cite{FeurerARMR2007,AshcroftSolidStatePhysics1976} $\EpsO(\omega) = \EpsO(\infty) + \left(
\EpsO(0) - \EpsO(\infty) \right) \OmgTO^2/\left(\OmgTO^2-\omega^2-\Ii\gamma_{\mathrm{o}}
\omega\right)$, $\EpsE(\omega) = \EpsE(\infty) + \left( \EpsE(0) - \EpsE(\infty) \right)
\OmgTE^2/\left(\OmgTE^2-\omega^2-\Ii\gamma_{\mathrm{e}} \omega\right)$, where $\EpsO(\infty)$,
$\EpsO(0)$, $\EpsE(\infty)$, $\EpsE(0)$ are the high-frequency and low-frequency limits of $\EpsO$
and $\EpsE$, $\gamma_{\mathrm{o}}$ and $\gamma_{\mathrm{e}}$ are the relaxation rates associated
with anharmonic optical-phonon decaying. We adopt their room-temperature ($\sim300$~K) values from
Ref.~\cite{FeurerARMR2007}, and set the low-temperature ($\sim100$~K) $\gamma_{\mathrm{o}}$ and
$\gamma_{\mathrm{e}}$ to be about one third of their room-temperature values according to
Ref.~\cite{IvanovaFerro1978,CapekPSSC2007,HuaMCP2002} $\EpsE$ and $\EpsO$ undergo near divergence
and sign change around $\OmgTO$ and $\OmgTE$, signifying the high-$Q$ resonant coupling between
photons and optical phonons. They turn back to positive at the longitudinal optical-phonon
frequencies $\OmgLO=\OmgTO \sqrt{\EpsO(0)/\EpsO(\infty)}=2\pi\times6.7$~THz and $ \OmgLE=\OmgTE
\sqrt{\EpsE(0)/\EpsE(\infty)}=2\pi\times12.3$~THz in this
model.\cite{FeurerARMR2007,AshcroftSolidStatePhysics1976}

\begin{figure}
\centerline{\includegraphics[scale=0.7]{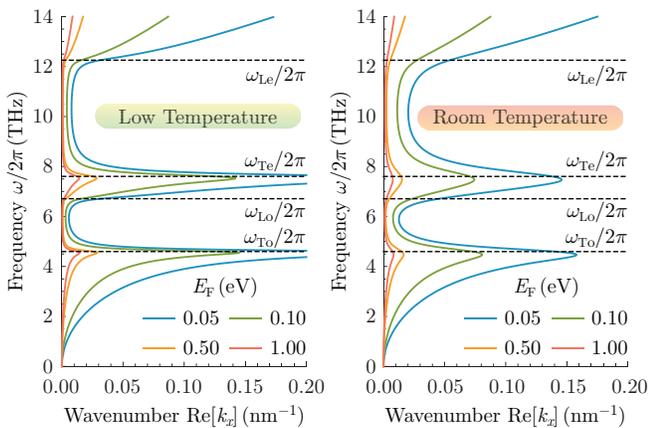}} \caption{Dispersion relations of surface
plasmon-phonon-polaritons for different Fermi energies on a graphene sheet clamped in LiNbO$_3$
crystals in the low-temperature ($\sim100$~K) and room-temperature ($\sim300$~K)
conditions.}\label{FigSpectrum}
\end{figure}

For the plane-wave study, we set $k_y=0$ and choose $k_x$ as the progressive wavenumber, which can
be explicitly solved as
\begin{equation}
k_x^2 = \EpsE(\omega)\frac{\omega^2}{c^2}  +
\frac{\EpsO(\omega)\EpsE(\omega)}{\left(2e^2|\EF|/\hbar^2\right)^2}
\omega^2\left(\omega+\Ii\frac{e\VF^2}{\mu|\EF|}\right)^2. \label{Dispersionkx}
\end{equation}
The modal wavelength $\lambda=2\pi/\RE[k_x]$, the attenuation length $\xi=1/\IM[k_x]$, and the
confining length $l = 1/\RE[\KapE]$ can be defined accordingly. For a Fermi energy between
0.05--1.0~eV, and driving frequency in the range of 1--10~THz, the dispersion relation is primarily
controlled by the second term of Eq.~(\ref{Dispersionkx}). The plasmon behavior is strongly
affected by the optical-phonon resonances. So the combined excitations are surface
plasmon-phonon-polaritons (SPPPS). Figure~\ref{FigSpectrum} shows the dispersion relation
$\omega(\RE[k_x])$ under different conditions. For each given $\EF$, the dispersion curves mainly
stay in the three allowed-bands: $\omega<\OmgTO$, $\OmgLO<\omega<\OmgTE$, and $\omega>\OmgLE$, but
weakly leak into the two forbidden-bands: $\OmgTO<\omega<\OmgLO$ and $\OmgTE<\omega<\OmgLE$.
Finite-valued relaxation rates broaden the sharp peaks around the optical-phonon resonances. For a
small $\EF$, the dispersion curves bend more considerably towards the optical-phonon lines and can
give very large $k_x$ at low frequencies; for a large $\EF$, the dispersion curves mostly attach to
the THz light line (very close to the $\omega$-axis and cannot be identified in the scale of
Fig.~\ref{FigSpectrum}). In reality, there exist other optical-phonon resonances higher than
$\OmgTO$ and $\OmgTE$,\cite{IvanovaFerro1978,CapekPSSC2007,HuaMCP2002} which will make the curves
in Fig.~\ref{FigSpectrum} more kinked than as shown. But we will only focus on the frequency region
close to the fundamental optical-phonon resonance $\OmgTO=2\pi\times4.6$~THz from below. In
Table~\ref{Results}, we list the calculated characteristic quantities at about 100~K. For low-power
sensitive THz-photon manipulation, a low temperature is helpful for suppressing the thermal noise
or dissipation.\cite{PeresRMP2010,DasSarmaRMP2011,IvanovaFerro1978,CapekPSSC2007,HuaMCP2002} In
Table~\ref{Results}, one can see clearly the huge effective refractive index and ultra-short modal
wavelength compared with the free-space wavelength of low-THz photons. Large dissipation occurs at
frequencies above 4.0~THz. But just below it, for $\EF\lesssim0.1$~eV, the wavelength can indeed be
squeezed to 100--200~nm while the confining length is only about 10--20~nm.

\begin{table}
\caption{Calculated effective refractive index $\tilde{n}$, modal wavelength $\lambda$, attenuation
length $\xi$, and confining length $l$ of the surface plasmon-phonon-polaritons with several
representative Fermi energies and frequencies at low temperature ($\sim 100$~K).}\label{Results}
\begin{tabular}{c|ccccc}
\hline
$\EF$~(eV) & $\omega/2\pi$~(THz) & $\tilde{n}$ & $\lambda$~(nm) & $\xi$~(nm) & $l$ (nm)\\
\hline
0.05 & 2.5 & 532+14\Ii & 225 & 1386 & 27 \\
     & 3.0 & 694+21\Ii & 144 & 760 & 16 \\
     & 3.5 & 922+37\Ii & 93 & 364 & 10 \\
     & 4.0 & 1335+95\Ii & 56 & 126 & 5 \\
     & 4.5 & 2836+1045\Ii & 23 & 10 & 1 \\
\hline
0.10 & 2.5 & 266+5.2\Ii & 450 & 3676 & 53 \\
     & 3.0 & 347+8.6\Ii & 288 & 1846 & 32 \\
     & 3.5 & 461+17\Ii & 186 & 820 & 19 \\
     & 4.0 & 668+45\Ii & 112 & 266 & 10 \\
     & 4.5 & 1420+518\Ii & 47 & 20 & 3 \\
\hline
0.50 & 2.5 & 53+0.8\Ii & 2252 & 24879 & 266 \\
     & 3.0 & 69+1.4\Ii & 1440 & 11133 & 162 \\
     & 3.5 & 92+3.0\Ii & 929 & 4563 & 96 \\
     & 4.0 & 134+8.5\Ii & 561 & 1397 & 49 \\
     & 4.5 & 284+103\Ii & 234 & 103 & 13 \\
\hline
1.00 & 2.5 & 27+0.4\Ii & 4504 & 52058 & 532 \\
     & 3.0 & 35+0.7\Ii & 2880 & 22856 & 324 \\
     & 3.5 & 46+1.5\Ii & 1858 & 9255 & 192 \\
     & 4.0 & 67+4.2\Ii & 1122 & 2812 & 97 \\
     & 4.5 & 142+51\Ii & 469 & 207 & 26 \\
\hline
\end{tabular}
\end{table}

As an example, we now employ the large difference in the length scale of SPPPs under different
Fermi energies to make low-power subwavelength waveguides. LiNbO$_3$ is known to possess very large
spontaneous polarization $P_{\mathrm{s}}$. In a bulk crystal under zero electric field,
$P_{\mathrm{s}}\approx 70$~$\mu$C~cm$^{-2}$,\cite{SannaPRB2010,MarekPhDThesis2003,JoshiAPL1993}
which is equivalent to a surface bound-charge density $n_{\mathrm{s}}\approx
4.4\times10^{14}$~cm$^{-2}$. In a thin film of about 200~nm thick,
$P_{\mathrm{s}}\approx5$~$\mu$C~cm$^{-2}$,\cite{JoshiAPL1993} and the equivalent surface
bound-charge density is $n_{\mathrm{s}}\approx 3.1\times10^{13}$~cm$^{-2}$. For our studies, we
take the 200~nm thickness for each slab of LiNbO$_3$. As shown in Fig.~\ref{FigDesign}, for the
parallel-poling configuration, the bound charges of opposite signs from the lower and upper slabs
cancel each other, leaving an approximately zero-potential setting for the graphene sheet. Thus the
charge-carrier density on graphene is solely determined by chemical doping. But for the
antiparallel-poling configuration, the bound charges of the same sign from the both slabs cause a
net positive or negative potential on the graphene sheet. For the 200~nm thin film clamping, the
induced charge-carrier density is $n_{\mathrm{c}}\approx 2n_{\mathrm{s}} =
6.2\times10^{13}$~cm$^{-2}$ which is equivalent to a nearly $\pm1$~eV electrostatic gating based on
the preceding calculation. In the numerical simulation below, we assume that a small $\EF=0.05$~eV
is built in graphene from chemical doping, and is reserved between the parallel-poling domains,
while a large $\EF\approx 0.05\pm1.0 \approx \pm 1.0$~eV is generated between the
antiparallel-poling domains due to the ferroelectric gating. The slight difference in $|\EF|$
between the electron and hole cases is neglected because of the smallness of 0.05~eV compared with
1.0~eV. The electron-hole symmetry in graphene plays an important role here.

\begin{figure}
\centerline{\includegraphics[scale=0.7]{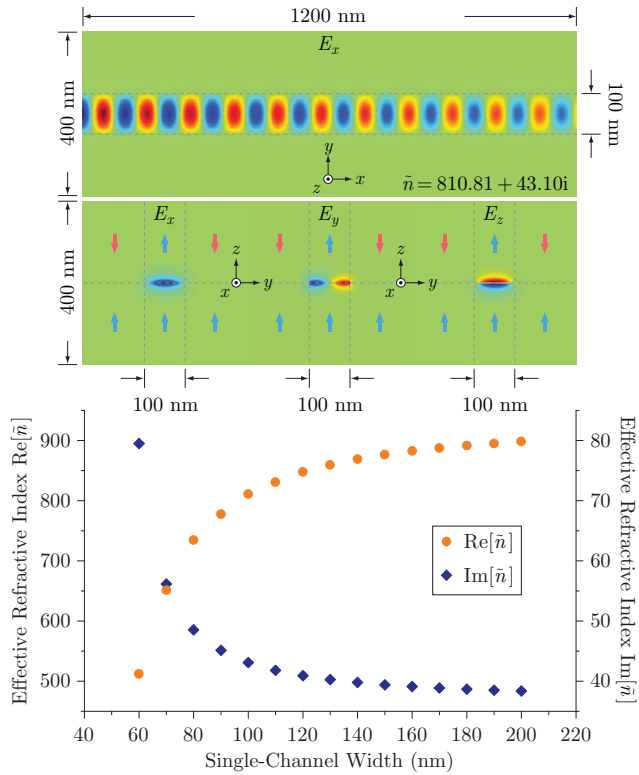}} \caption{Simulated effective
refractive index and electric-field profile of the fundamental waveguide mode in a parallel-poling
channel between two antiparallel-poling barriers. (Blue and red arrows indicate domain
orientations.)}\label{FigSingleChannel}
\end{figure}

We choose the frequency 3.5~THz for all our finite-difference time-domain (FDTD) simulations. The
free-space photon wavelength is 85.7~$\mu$m. For the domain poling configuration shown in
Fig.~\ref{FigSingleChannel}, a single $\EF=0.05$~eV channel is produced in between two $\EF=1.0$~eV
barriers. As can be inferred from our previous discussion, a low-$\EF$ channel is more
``dielectric" in the sense that it carries less charges but hosts more photons, whereas a
high-$\EF$ channel is more ``metallic" in the sense that it carries more charges but permits less
photons. Thus the deep subwavelength SPPP modes preferably flow in the middle channel with very
tiny penetration into the left- and right-barriers. The lower panel of Fig.~\ref{FigSingleChannel}
shows the real and imaginary parts of the effective refractive index $\tilde{n}$ changing with the
channel width. One can see that the waveguide mode undergoes exponentially stronger attenuation and
weaker subwavelength after the channel width goes down to below 100~nm, consistent with the 93~nm
plane-wave modal wavelength at 3.5~THz in Table~\ref{Results}. We choose 100~nm to plot the
profiles of each electric-field component in the upper panel of Fig.~\ref{FigSingleChannel}. For
the effective refractive index $\tilde{n}=810.81+43.10\Ii$ in this case, the modal wavelength
$\lambda=105.6$~nm, and the attenuation length $\xi=316.2$~nm.

\begin{figure}
\centerline{\includegraphics[scale=0.7]{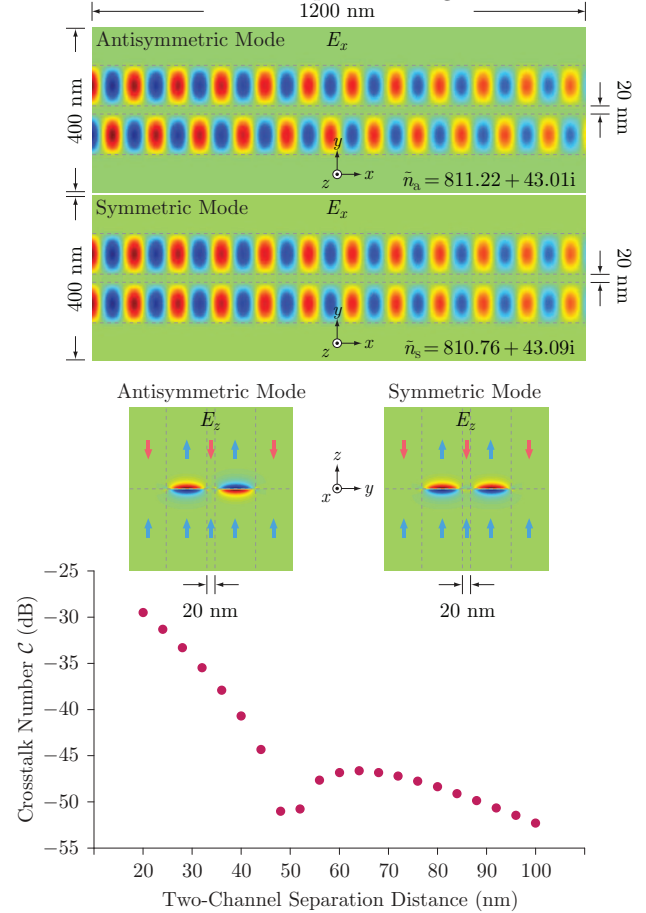}} \caption{Simulated crosstalk number and
electric-field profiles of the antisymmetric and symmetric waveguide modes in two 100~nm wide
parallel-poling channels. (Blue and red arrows indicate domain orientations.)}\label{FigTwoChannel}
\end{figure}

The exceptional confining quality of the prescribed ferroelectric-gated channel can be manifested
by putting two such channels close to each other and see how the modes in the two channels become
coupled by varying the separation
distance.\cite{ChristensenACSNano2012,YehDielectricWaveguides2008,VeronisOE2008} We may define a
dimensionless crosstalk number,
\begin{equation}
\mathcal{C}\equiv
-10\log_{10}\left|\frac{\mathrm{Re}[\tilde{n}_{\mathrm{a}}]+\mathrm{Re}[\tilde{n}_{\mathrm{s}}]}
{4(\mathrm{Re}[\tilde{n}_{\mathrm{a}}]-\mathrm{Re}[\tilde{n}_{\mathrm{s}}])}\right|,
\end{equation}
which estimates (in terms of dB) the number of modal (not free-space) wavelength needed, for an
injected power initially in one channel to be transferred into the other and then be transferred
back. In Fig.~\ref{FigTwoChannel}, we can see the simulated crosstalk number to be mostly in the
range of $-45$ to $-50$~dB. For the plotted pattern of 20~nm separation distance, this number is
still far below zero at $-29.5$~dB. The dissipation has certainly come in at a much shorter
propagation distance, so the waves cannot really travel that far.\cite{VeronisOE2008} But this
number still shows the extreme confining quality in these ferroelectric-graphene waveguides
compared with conventional dielectric and plasmonic waveguides. For example, silicon waveguides
with similar dimensions at infrared frequencies have a crosstalk number of the order of $-10$~dB in
accordance with our definition.\cite{YehDielectricWaveguides2008} Ba$_{0.5}$Sr$_{0.5}$TiO$_3$-metal
inter-layer plasmonic waveguides at visible-light frequencies can have a crosstalk number of the
order of $-30$~dB, but must take a much larger separation distance.\cite{LiuAPL2006} One may notice
a singular drop at about 50~nm on the calculated curve in the lower panel of
Fig.~\ref{FigTwoChannel}. Based on an analysis to the dipole-dipole coupling in this particular
system, we find it to be due to a competition between the relative magnitudes of field components.
A large $E_y$ field tends to make the symmetric mode have a higher refractive index, while a large
$E_x$ or $E_z$ field tends to make the antisymmetric mode have a higher refractive index. The 50~nm
separation distance between two 100~nm wide waveguides happens to be the turning point, where
$\mathrm{Re}[\tilde{n}_{\mathrm{a}}]-\mathrm{Re}[\tilde{n}_{\mathrm{s}}]\simeq 0$. A more thorough
study on this phenomena will be performed.

We acknowledge the financial support by NSF (ECCS Award No. 1028568) and the AFOSR MURI (Award No.
FA9550-12-1-0488).

\bibliography{RefsShort}

\end{document}